\documentclass[conference]{IEEEtran}
\IEEEoverridecommandlockouts
\usepackage{cite}
\usepackage{amsmath,amssymb,amsfonts}
\usepackage{algorithm}
\usepackage{algpseudocode}
\usepackage{graphicx}
\usepackage{textcomp}
\usepackage{xcolor}
\usepackage{booktabs} 
\usepackage[hyphens]{url}
\usepackage[hidelinks]{hyperref}
\usepackage{comment}

\usepackage{array}
\usepackage{ragged2e}
\usepackage{subcaption}
\def\BibTeX{{\rm B\kern-.05em{\sc i\kern-.025em b}\kern-.08em
    T\kern-.1667em\lower.7ex\hbox{E}\kern-.125emX}}

\begin{document}

    \title{SWARM+: Scalable and Resilient Multi-Agent Consensus for Decentralized Data-Aware Workload Management}

    \author{\IEEEauthorblockN{
    Komal Thareja\IEEEauthorrefmark{1},
    Krishnan Raghavan\IEEEauthorrefmark{2},
    Hamza Safri\IEEEauthorrefmark{3},
    Anirban Mandal\IEEEauthorrefmark{1},
    Ewa Deelman\IEEEauthorrefmark{3}
    }
    
    \IEEEauthorblockA{\IEEEauthorrefmark{1}Renaissance Computing Institute, University of North Carolina at Chapel Hill, NC, USA}
    
    \IEEEauthorblockA{\IEEEauthorrefmark{2}Argonne National Laboratory, Lemont, IL, USA}
    
    \IEEEauthorblockA{\IEEEauthorrefmark{3}Information Sciences Institute, University of Southern California, Marina del Rey, CA, USA}
    }
\maketitle

\begin{abstract}
Distributed scientific workflows are increasingly executed across heterogeneous and geo-distributed computing environments, where centralized workload orchestration becomes a scalability and resilience bottleneck. This paper presents \textbf{SWARM+}, a decentralized workload management system that coordinates workload placement through hierarchical multi-agent consensus, reducing coordination overhead and dramatically improving scalability, while tolerating failures and dynamic membership changes. \textbf{SWARM+} enables data-aware scheduling policies that incorporate resource availability, data transfer node (DTN) connectivity, and data locality into workload placement decisions. We evaluate SWARM+ on the distributed FABRIC testbed using heterogeneous scientific workloads derived from production workflow traces obtained from the Pegasus Workflow Management System (WMS). Experimental results show that SWARM+ scales coordination to 990 distributed agents with approximately 1\,s per-job selection time at 110 agents. SWARM+ demonstrates balanced workload distribution, maintains over $97\%$ job completion under distributed failures with graceful degradation (mean ${\sim}95\%$ job completion) during correlated site outages, tolerates coordinator agent failures gracefully, improves schedule quality by employing data-aware policies, and reduces both selection time and scheduling latency by $97$--$98\%$ when compared to the prior SWARM system.
\end{abstract}


\section{Introduction}

Modern scientific workflows process massive amounts of data from diverse instruments and sensors across geographically distributed, heterogeneous compute and storage systems, from leadership-class supercomputers to edge systems, all connected by high-performance networks. This heterogeneity creates resilience challenges across the stack, spanning applications, workload management, filesystems, storage, networks, and hardware. 

While the complexity of the applications and their execution environment has grown, workload management has remained largely centralized. Most workflow management systems (WMSs)~\cite{deelman2020jocs, galaxy2022galaxy, nextflow} are vertically integrated and centralized, providing submission and orchestration functions but introducing scalability bottlenecks and single points of failure. Centralized schedulers and resource managers such as SLURM~\cite{slurm}, PBS~\cite{pbs}, and HTCondor~\cite{condor} share similar vulnerabilities, where controller failures can halt workloads and degrade performance at scale.

Resilience strategies for scientific workloads often require expert-designed, static mechanisms and rely on SLA management~\cite{central_aminvahdat, 7523331}. Prior fault-tolerant workflow approaches have focused on specific components—such as failure detectors, cross-facility MPI, or logging-based architectures~\cite{FT-GRID3, FT-GRID1, FT-GRID4}—while assuming a reliable central manager. These centralized approaches do not scale to modern scientific infrastructures with thousands of sensors and hundreds of distributed compute and storage resources, especially under transient or hard cyberinfrastructure failures.

These challenges motivate a {\em radically different, fully decentralized approach} to scalable and resilient scientific workload management. Our solution leverages multi-agent systems (MAS) and adapts them to the workload management problem. 


Our prior work, SWARM~\cite{ccgrid25}, focused on a key subproblem in that vision: the {\em distributed job selection problem}, where globally distributed, heterogeneous resource agents must autonomously choose workloads from a dynamic distributed global workload/job pool—without any central orchestration. We demonstrated how distributed resource agents can coordinate workload selection by leveraging a decentralized greedy consensus mechanism employing a peer-to-peer three-phase protocol inspired by Practical Byzantine Fault Tolerance (PBFT)~\cite{pbft}. Although the approach performed 63.5\% better than the vanilla PBFT, the flat mesh topology in SWARM exhibited $O(n^2)$ communication complexity resulting in poor scalability as the number of participating agents grew. In addition, it provided no explicit failure recovery, lacked dynamic elastic agent scaling, and did not take into consideration data locality in its cost model. The latter is important when considering many of the current workloads, which are data intensive (e.g. instrument data, AI model training). 

To address these limitations, this paper presents \textbf{SWARM+}, a scalable and resilient decentralized workload management system for geo-distributed scientific infrastructures. SWARM+ extends the original SWARM design through three novel contributions:
(1)~a hierarchical coordination architecture that reduces global coordination overhead across distributed agents resulting in efficient consensus and improved scalability; 
(2)~decentralized resilience mechanisms that tolerate a variety of agent failures and dynamic membership changes; and
(3)~data-aware scheduling policies that incorporate resource feasibility checks, improved cost models, network connectivity, and DTN locality.

We evaluate SWARM+ on the FABRIC testbed~\cite{fabric}, a programmable network research infrastructure spanning more than 30 geographically distributed sites, with a variety of agent counts, agent topologies, real-world workloads, and covering a range of failure scenarios and network conditions. 

\section{System Architecture}
\label{sec:architecture}

\begin{figure*}[t]
\centering

\begin{subfigure}[t]{0.55\textwidth}
\centering
\includegraphics[width=\textwidth]{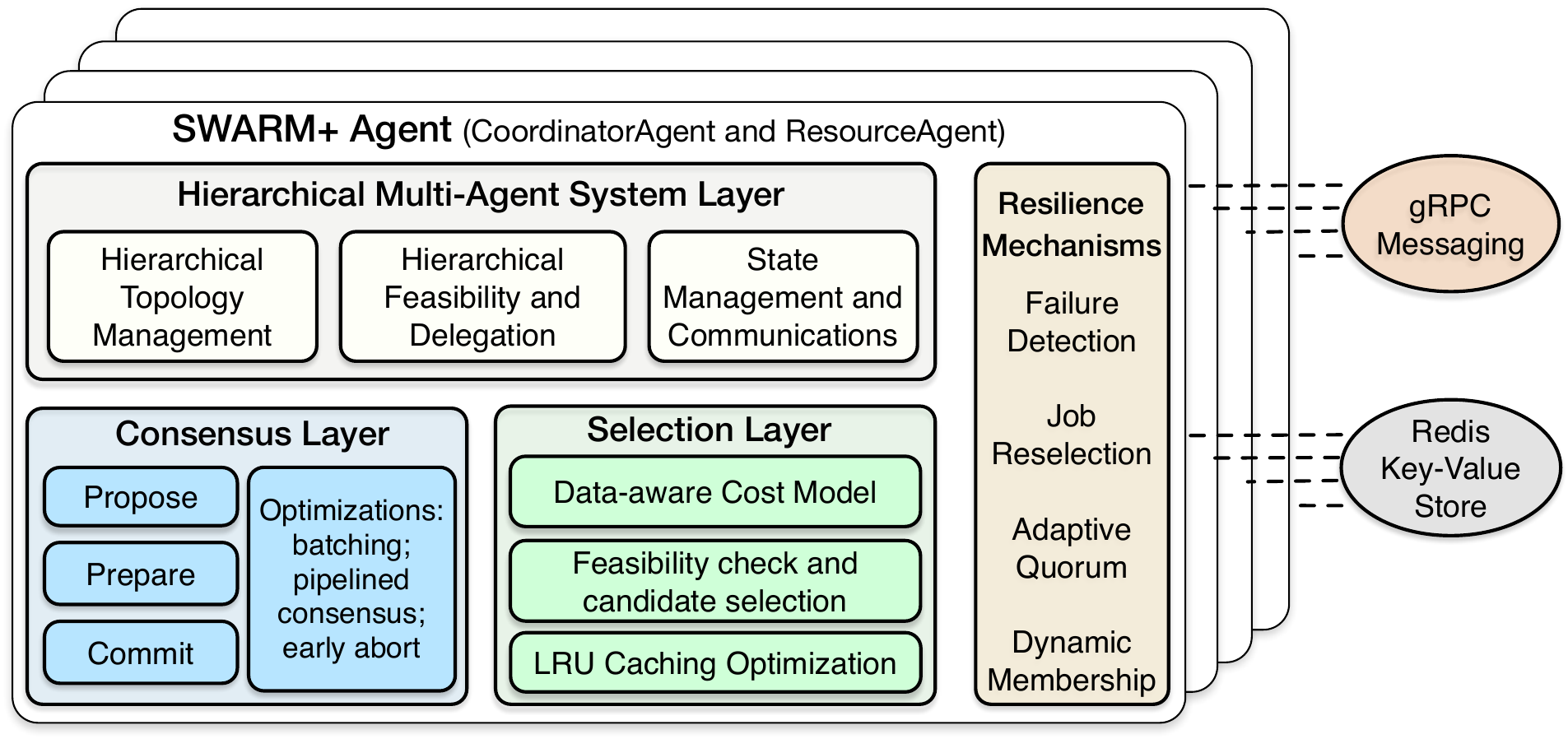}
\caption{SWARM+ architecture.}
\label{fig:architecture}
\end{subfigure}
\hfill
\begin{subfigure}[t]{0.4\textwidth}
\centering
\includegraphics[width=\textwidth]{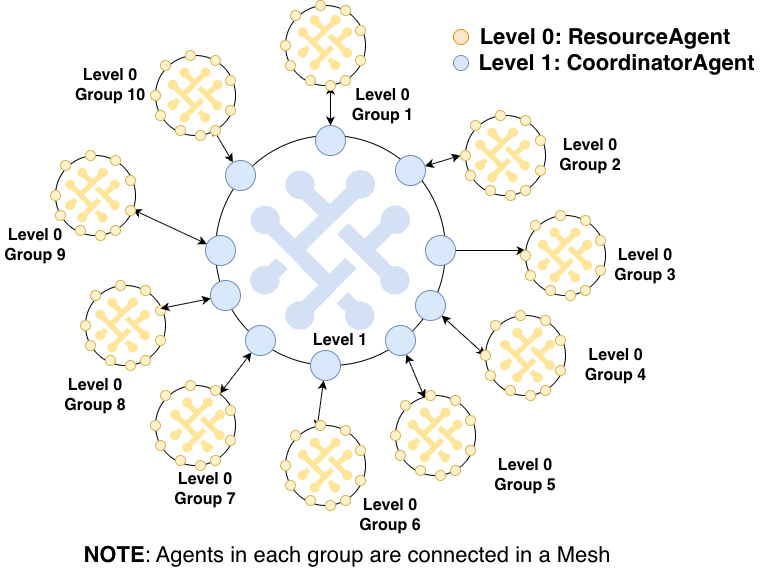}
\caption{Hierarchical-110 deployment topology.}
\label{fig:hier-topology}
\end{subfigure}
\vspace{-4pt}
\caption{SWARM+ system: hierarchical coordination and deployment organization.}
\vspace{-14pt}
\label{fig:system-overview}
\end{figure*}

The SWARM+ system (Figure~\ref{fig:system-overview}) consists of three closely interconnected 
layers: (1) the {\bf Hierarchical Multi-Agent System Layer} that is responsible for hierarchical coordination, agent topology management, hierarchical feasibility and cross layer delegation, and overall state management, (2) the {\bf Consensus Layer} that manages and optimizes consensus across the hierarchical multi-agent system, and (3) the {\bf Selection Layer} performing distributed job selection using feasibility filtering, cost-based ranking and data-aware policies. Resilience mechanisms are built into each layer of the system.

\subsection{Hierarchical Multi-Agent System Layer}
\label{subsec:topology}
\begin{figure}[h]
\centering
\includegraphics[width=0.4\textwidth]{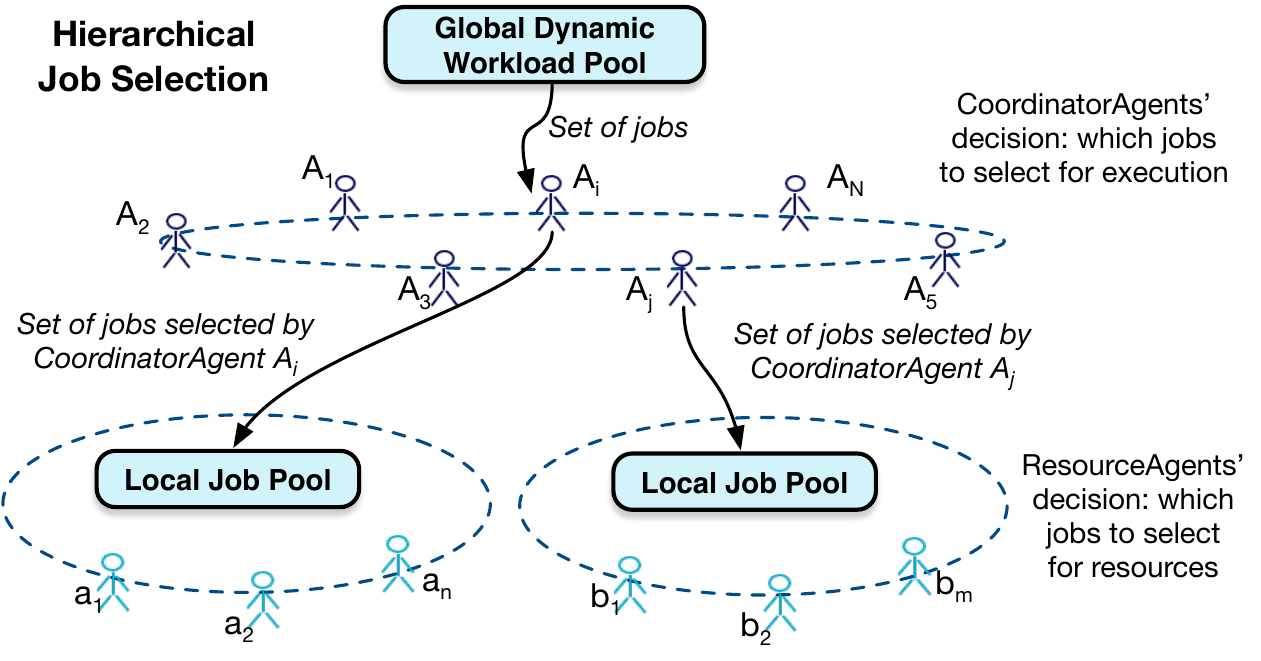}
\caption{Hierarchical job selection (2-level scenario).}
\label{fig:job-selection}
\end{figure}

The SWARM+ system organizes distributed agents into a 
configurable-depth, hierarchical tree structure with distinct coordination roles. 
Figure~\ref{fig:job-selection} illustrates this model in a two-level deployment. 
At the higher level, coordinator agents (CoordinatorAgents) select jobs from a shared global workload pool and delegate them to groups of resource agents (ResourceAgents). Within each group, at the lower level, ResourceAgents coordinate locally to determine workload placement and allocation.
This organization enables scalable coordination through group-wise aggregation and cross-layer delegation. For a two-level organization, we have:

\begin{itemize}

\item \textbf{Level~0 (ResourceAgents):}
These agents maintain local resource state including CPU, RAM, GPU, disk, and DTN connectivity information. ResourceAgents select jobs from local job pools, coordinate resource allocation within their group, and execute workloads on local resources. Resource agents are grouped according to physical deployment sites, allowing most coordination to remain local.

\item \textbf{Level~1 (CoordinatorAgents):}
These agents coordinate workload delegation across groups of ResourceAgents. CoordinatorAgents aggregate group-level state information, participate in higher-level consensus, and delegate workloads to appropriate resource-agent groups. CoordinatorAgents do not directly execute workloads.
\end{itemize}

The hierarchy supports both horizontal and vertical scaling. Horizontal scaling is achieved by adding groups at the same hierarchy level, while vertical scaling is achieved by introducing additional coordination levels as the infrastructure grows.

\subsubsection{Hierarchical Topology Management: Group Formation and Agent Discovery}
 Group assignment is determined during deployment through a configuration generation phase. Each agent receives a \emph{topology descriptor} specifying hierarchy level, group identifier, peer set, and parent reference. Groups are aligned with physical sites; agents deployed at the same site belong to the same Level~0 group with Level~1 CoordinatorAgents assigned to each group.

At runtime, agents identify active peers through periodic heartbeat-based discovery mechanisms using a Redis~\cite{redis} key-value store. Each agent publishes its state, including host, port, capacities, and load, to a group-scoped Redis namespace, while peers poll this namespace during refresh cycles. This design avoids runtime group negotiation while supporting elastic scaling and dynamic agent addition. Newly started agents load their pre-assigned topology configuration and discover active peers through Redis.

\subsubsection{Hierarchical Delegation}
Hierarchical delegation is organized through parent-child coordination relationships across multiple hierarchy levels. Lower-level agents operate as children, higher-level agents operate as parents, and same-level agents act as peers for consensus communication. Agents maintain local topology metadata including hierarchy level, group identifier, parent reference, child set, and peer set.

CoordinatorAgents delegate workloads to groups of ResourceAgents rather than coordinating directly with every resource in the system. ResourceAgents perform local workload placement and execution using exact local resource information, while higher-level agents coordinate workload delegation using summarized and aggregated group-level information.

\subsubsection{Aggregation-Based Feasibility Checking}
If all parent agents queried every child directly during consensus, communication overhead would scale as $O(n \cdot m)$, where $n$ is the number of parent agents and $m$ is the maximum number of children per parent. To reduce coordination overhead, SWARM+ uses aggregation functions that summarize child-group capabilities.

\begin{itemize}
\item \textbf{Maximum Child Capacity:}
Each parent aggregates child resource availability into a capacity vector

\vspace{-10px}

\[
\mathbf{C}_{\max} = \max_{c \in \text{children}} c.\text{capacity},
\]

where $c.\text{capacity}$ represents available capacity across resources such as CPU, RAM, GPU, and storage.

\item \textbf{Aggregated DTN Connectivity:}
Parents aggregate DTN connectivity into a union of reachable DTN endpoints with associated connectivity scores summarizing bandwidth, latency, and reliability.

\end{itemize}

Using these summaries, parents perform $O(1)$ feasibility checks per workload to determine whether a child group can potentially satisfy resource requirements, avoiding $O(m)$ per-child queries during consensus.
The aggregation strategy affects only the CoordinatorAgent feasibility filter; final workload placement is determined by Level~0 intra-group consensus using exact per-agent costs.

\subsubsection{Delegation Monitoring}
SWARM+ employs delegation monitoring to handle cases where delegated workloads cannot be successfully placed or completed within a target child group. Each parent associates a \emph{delegation timeout}. Workloads remaining in delegated-but-incomplete state beyond the timeout are treated as failed and reassigned to the parent-associated job pool with exclusion flags preventing immediate reassignment to the same child group. This approach trades perfect placement accuracy, which would require expensive child-level queries during scheduling, for scalable distributed delegation combined with monitoring-driven recovery.

\subsubsection{State Management and Communication}
Agents coordinate through distributed state management and inter-agent messaging infrastructure. State associated with resources, job pools, and consensus metadata, including proposal identifiers, live-neighbor sets, and job states, must be shared across distributed agents to support workload coordination and consensus operations. The state is maintained through a distributed Redis key--value store configured with replication for resilience. Structured namespaces support scoped coordination within hierarchy levels and agent groups. Inter-agent messaging uses low-latency gRPC~\cite{grpc} communication with bidirectional streaming channels between agent pairs; reducing connection overhead during coordination and consensus operations. If a communication channel disconnects, clients automatically reconnect.

\subsection{Consensus Layer}
\label{subsec:consensus}
\vspace{-3px}
Consensus in SWARM+ is organized hierarchically to localize coordination and reduce communication overhead across geographically distributed agents. Rather than requiring all agents to participate in global consensus, consensus operations are performed both across coordinator groups and within local resource-agent groups.

\paragraph{Fault Model}
SWARM+ targets crash fault tolerance: agents are trusted but may fail by stopping (process crashes, network partitions, or host failures). Byzantine behavior (malicious or arbitrarily incorrect agents) is outside the current threat model, as agents operate within trusted administrative domains. The majority quorum $\lceil(n_{\text{live}}+1)/2\rceil$ guarantees safety under crash faults, and the three-phase protocol structure ensures that no two agents commit conflicting assignments for the same job.

\paragraph{Consensus Protocol}
Each agent, either a CoordinatorAgent or ResourceAgent, participates in a three-phase consensus protocol inspired by PBFT~\cite{pbft}.

\textbf{Proposal Phase:}
An agent $a_i$ wishing to execute job $j$ broadcasts
\[
\langle \text{PROPOSAL}, p, j, a_i, c \rangle,
\]
where $p$ is a unique proposal identifier and $c$ is the computed scheduling cost.

\textbf{Prepare Phase:}
Upon receiving a proposal, agent $a_k$ validates feasibility and detects conflicts. If no better proposal exists for job $j$, the agent broadcasts
\[
\langle \text{PREPARE}, p, j, a_i, a_k \rangle.
\]

Prepare messages accumulate until reaching quorum count
\[
q =
\left\lceil
\frac{n_{\text{live}}+1}{2}
\right\rceil,
\]

where $n_{\text{live}}$ denotes the number of currently live agents.

\textbf{Commit Phase:}
After reaching prepare quorum, agents broadcast
\[
\langle \text{COMMIT}, p, j, a_i, a_k \rangle
\]

and wait for commit quorum before finalizing workload selection.

\paragraph{Hierarchical Consensus Execution}

Each workload undergoes two sequential consensus rounds. During \emph{inter-group consensus}, CoordinatorAgents execute the protocol to determine job-to-group assignment. The selected coordinator then delegates the workload to the target child group through the distributed job pool.

During \emph{intra-group consensus}, ResourceAgents within the selected group execute the same protocol to determine workload placement on a specific resource. Consensus messages remain confined within the target group, allowing most coordination to remain local.


For $N$ agents organized into $G$ groups of size $g = N/G$, flat three-phase consensus requires $O(N^2)$ messages per job. In contrast, hierarchical consensus in SWARM+ requires 
$O(g^2 + G^2) = O((N/G)^2 + G^2)$
messages per job. When $G=\sqrt{N}$, communication complexity reduces to approximately $O(N)$.
For a hierarchy with $L$ levels and constant group size $g$, the per-job coordination complexity becomes
$O(g^2 \cdot L)$,
which approaches $O(\log N)$ when $L=\log_g N$.

\paragraph{Consensus Optimizations}

SWARM+ incorporates several optimizations to improve consensus throughput: \textbf{(1) Proposal Batching:}
Agents batch up to $B$ jobs per consensus round, amortizing message overhead across multiple workload selections; \textbf{(2) Pipelined Consensus:}
Multiple consensus rounds may proceed concurrently using unique proposal identifiers while maintaining efficient conflict detection; \textbf{(3) Early Abort on Conflicts:}
When receiving a proposal for job $j$, agents compare proposals using cost-based ordering. Better proposals trigger immediate rejection, allowing proposers to abort and reselect workloads without timeout expiration.

\subsection{Selection Layer}
\label{subsec:selection}
\vspace{-3px}
The selection layer performs distributed job selection using feasibility filtering, cost-based ranking, and data-aware scheduling policies. While the consensus layer coordinates agreement across agents, the selection layer determines which workloads' agents should propose during consensus. Job selection proceeds through feasibility checking, cost computation, and candidate selection.

\subsubsection{Feasibility Checking}
Before computing scheduling costs, agents evaluate whether a workload can execute on a target resource. For an agent--job pair $(a,j)$, feasibility is defined as
\[
\text{feasible}(a,j)
=
\bigwedge_{r \in R}
(a.r_{\text{available}} \geq j.r_{\text{req}})
\land
\text{conn\_feasible}(a,j),
\]

where $R=\{\text{CPU},\text{RAM},\text{Disk},\text{GPU}\}$. Connectivity feasibility (conn\_feasible) ensures that the DTN requirements can be satisfied either through direct access or acceptable cross-site transfer paths. Infeasible agent--job pairs receive infinite cost and are excluded from further consideration.

\subsubsection{Cost Model}

For each feasible agent--job pair $(a,j)$, scheduling cost combines resource utilization and workload penalties:
\[
\text{cost}(a,j)
=
\sum_{r \in R} w_r \cdot u_r(a,j)
+
\sum_{p \in P} \text{penalty}_p(a,j):\] 
where \[
 u_r(a,j)=\frac{j.r_{\text{req}}}{a.r_{\text{available}}}
\]
is the utilization fraction and $P=\{\text{connectivity},\text{long\_job}\}$ denotes penalty types. Resource weights ($w_r$) adapt to workload characteristics; compute-intensive workloads prioritize CPU/GPU resources, memory-intensive workloads emphasize RAM, and data-intensive workloads prioritize storage and connectivity.

\paragraph{Connectivity Penalty}

Data locality is incorporated through DTN connectivity penalties. Jobs specify input and output data requirements through DTN endpoint metadata, allowing the system to identify the required DTN set
\[
D_j
=
\{d \mid d \in j.\text{data\_in} \cup j.\text{data\_out}\}.
\]

Agents maintain DTN connectivity profiles containing quality-of-service scores in the range $[0,1]$. Average connectivity is computed as
\[
\overline{s}
=
\frac{1}{|D_j|}
\sum_{d \in D_j} s_d,
\]

and connectivity penalty is defined as
\[
P_{\text{conn}}
=
1 + \beta(1-\overline{s}),
\]

where $\beta$ controls the influence of data locality during scheduling.

\paragraph{Long-Job Penalty}

Jobs exceeding a configurable duration threshold incur additional scheduling penalty of
$\alpha \cdot (j.\text{walltime}-\tau)$,
where $\tau$ is the duration threshold and $\alpha = 1/\tau$ is a normalization factor.

\subsubsection{Candidate Selection}

After cost computation, each agent constructs a cost matrix where rows represent agents and columns represent jobs. For each workload, an agent identifies itself as a candidate if its scheduling cost falls within threshold $\theta$ of the minimum observed cost across all agents. Candidate agent--job pairs are then forwarded to the consensus layer.

\subsubsection{Caching Optimization}
To reduce repeated feasibility and cost computations during steady-state execution, SWARM+ employs version-based caching. Cache keys are constructed from 
$(a.\text{id}, j.\text{id}, a.\text{version}, j.\text{version})$,
where agent and job version identifiers are incremented during state changes. Caching follows Least Recently Used (LRU) eviction and short Time To Live (TTL) policies, significantly reducing repeated cost computations and workload selection latency.

\subsection{Resilience and Elasticity}
\label{subsec:resilience}

SWARM+ incorporates decentralized resilience mechanisms that allow workload coordination to continue under failures, dynamic resource changes, and elastic agent scaling. This is achieved through multi-signal failure detection, automatic workload reselection, adaptive quorum management, dynamic membership support, and coordinator failover mechanisms.

\subsubsection{Failure Detection}

Agents detect failures using multiple runtime signals. Communication failures are identified through gRPC health checks, while Redis heartbeat expiry mechanisms identify agents that fail to refresh their state within configured timeout windows. Stale agents are removed from neighbor maps once delegation timeouts expire. Combining communication-level and heartbeat-based detection improves resilience against transient network failures and partial communication outages.

\subsubsection{Job Reselection}

When agent $a_f$ is detected as failed, surviving agents automatically reassign workloads previously associated with that agent. Workloads mapped to $a_f$ are reset to pending state and re-enter the selection pipeline. To prevent deadlock caused by failed proposers, workloads remaining in prepare or commit states beyond timeout thresholds are also reset and returned to selection.

\subsubsection{Adaptive Quorum}

SWARM+ dynamically adapts quorum size without requiring explicit reconfiguration. Each group computes quorum at time $t$ as
\[
q(t)
=
\left\lceil
\frac{n_{\text{live}}(t)+1}{2}
\right\rceil,
\]

where $n_{\text{live}}(t)$ denotes the number of responsive agents inferred from Redis-backed heartbeats.

As agents fail or become unreachable, quorum decreases proportionally, allowing consensus to continue under reduced membership. When agents join, quorum automatically expands to reflect the updated number of active participants.

\subsubsection{Dynamic Membership}

SWARM+ supports elastic agent addition across all hierarchy levels. Newly started agents register through Redis and announce availability through gRPC communication channels. Existing agents discover new participants during neighbor refresh cycles and incorporate them into future consensus rounds.

\subsubsection{Coordinator Failover}

CoordinatorAgent failures require additional handling because failed coordinators may interrupt workload delegation for entire child groups. To avoid orphaned groups, SWARM+ employs a co-parent shared-parenting mechanism in which each Level~0 group is assigned $K$ co-parents selected from the Level~1 coordinator pool.

Only the active leader, defined as the lowest-identifier live co-parent, delegates workloads to the child group. Remaining co-parents operate as warm standbys that continuously refresh child-group state and rapidly assume leadership after failure.

Leader election is deterministic and local, requiring no distributed locks or additional coordination rounds. When the active leader fails, the next eligible co-parent automatically assumes delegation responsibility after heartbeat expiration.

This mechanism reduces the probability of orphaned child groups from $p$ to $p^K$, where $p$ is the probability of individual coordinator failure.

\section{Evaluation}
\label{sec:evaluation}

We evaluate SWARM+ through real-world experimentation on the FABRIC testbed~\cite{fabric}, a programmable network research infrastructure designed to facilitate advanced experiments in networking and distributed computing. All SWARM+ source code is publicly available on GitHub~\cite{swarmagents}, and all evaluation data and scripts are released for reproducibility~\cite{swarmdata}.

We evaluate SWARM+ across four dimensions: (1)~coordination performance, measured by selection time and scheduling latency, relative to prior SWARM approach~\cite{ccgrid25}; 
(2)~scalability under increasing agent counts and heterogeneous communication topologies; (3)~resilience under distributed failures, correlated site outages, and coordinator failures; and (4)~sensitivity to data-aware scheduling and network impairment.

\subsection{Experimental Setup}
\label{subsec:eval-setup}

\textbf{Infrastructure and Topology.}
All experiments are run on the FABRIC testbed using 30~VMs (8~vCPUs, 16\,GB RAM each) distributed across three sites: Georgia Tech (GATECH), Clemson (CLEM), and MAX, with 10~VMs per site. When experiments require more than 30 agents, multiple agents execute concurrently on a single VM.

We evaluate two communication topologies:
(a)~\emph{Mesh}, a single-level fully connected topology used as a baseline for small-scale deployments; and
(b)~\emph{Hierarchical}, composed of Level~0 groups coordinated by Level~1 agents, with Level~2 coordinators introduced for larger-scale deployments, as described in \S\ref{subsec:topology}. Level~0 agents form local meshes within each group.

\textbf{Workloads and Agent Profiles.}
Table~\ref{tab:profiles} summarizes the profiles of the heterogeneous resources that agents are managing, which are used throughout the evaluation. Profiles are randomly assigned and proportionally distributed across sites.

\begin{table}[h]
\centering
\small
\caption{Agent resource profiles.}
\begin{tabular}{lccccc}
\toprule
\textbf{Profile} & \textbf{Ratio} & \textbf{CPU} & \textbf{RAM} & \textbf{Storage} & \textbf{GPU} \\
\midrule
Small & 40\% & 2 & 8\,GB & 100\,GB & 0 \\
Medium & 25\% & 4 & 16\,GB & 250\,GB & 0 \\
Large & 35\% & 8 & 32\,GB & 500\,GB & 4 \\
\bottomrule
\end{tabular}

\label{tab:profiles}
\end{table}

Unless otherwise noted, experiments use 2188 jobs derived from 10 production Pegasus~\cite{deelman2020jocs} workflow traces spanning diverse scientific domains including air quality modeling, genome-wide association studies, metagenome assembly, protein structure prediction, and RNA sequencing.

The workloads exhibit heterogeneous compute, memory, storage, and GPU requirements ranging from lightweight CPU-only tasks to resource-intensive GPU workloads. 

We additionally use ten DTN endpoints with connectivity scores in $[0.6,0.95]$. Agents are associated with 1--4 DTNs, and jobs inherit DTN dependencies from their originating agent profile, creating implicit data-locality preferences during scheduling.

\textbf{Execution Protocol and Evaluation Metrics.}
Each experiment initializes agents; submits workloads at controlled rates; executes continuously with runtime logging enabled; and exports metrics for offline analysis. Configurations are repeated between 5 and 50 times depending on workload size and topology configuration. We evaluate decentralized coordination efficiency using selection time and end-to-end scheduling latency,
\[
L_{\text{sched}}
=
L_{\text{wait}}
+
L_{\text{select}},
\]
where $L_{\text{wait}}$ denotes queue wait time, i.e. time between job submission and start of job selection, and $L_{\text{select}}$ denotes consensus-based selection latency, i.e. time between start of job selection and finalization of job assignment. Load balance is evaluated using Jain’s Fairness Index,
\[
J
=
\frac{
\left(\sum_{i=1}^{n} x_i\right)^2
}{
n \cdot \sum_{i=1}^{n} x_i^2
}
\]
where $x_i$ denotes the number of jobs assigned to agent $i$. The evaluation metrics used throughout the experiments are summarized in Table~\ref{tab:metrics}.

\begin{table}[h]
\centering
\footnotesize
\caption{Primary evaluation metrics.}
\label{tab:metrics}
\setlength{\tabcolsep}{4pt}
\begin{tabular}{lp{4.8cm}}
\toprule
\textbf{Metric} & \textbf{Description} \\
\midrule
Selection Time & Consensus coordination overhead. \\
Scheduling Latency & End-to-end scheduling delay. \\
P95 Selection & Tail coordination latency. \\
Throughput & Sustained scheduling capacity. \\
Makespan & Total time to complete all jobs. \\
Completion Rate & \% of successfully completed jobs. \\
Jain's Fairness & Workload distribution fairness. \\
WAN Slowdown & Multi-site latency increase. \\
\bottomrule
\end{tabular}

\end{table}

Unless otherwise noted, statistical comparisons use two-tailed $T$-tests ($\alpha = 0.05$) together with Cohen’s $d$ effect size measurements.

\subsection{Performance Evaluation}
\label{subsec:performance}

\textbf{Comparison with SWARM.}
To quantify the impact of the architectural improvements introduced by SWARM+, we reproduce the original SWARM~\cite{ccgrid25} configuration (Mesh-10, 100 synthetic jobs, multi-site). Table~\ref{tab:ccgrid-comparison} reports results averaged over 100 runs for SWARM and 50 runs for SWARM+.

\begin{table}[h]
\centering

\caption{SWARM+ vs.\ SWARM (Mesh-10, 100 jobs, multi-site)}
\label{tab:ccgrid-comparison}
\footnotesize
\setlength{\tabcolsep}{4pt}
\begin{tabular}{lccc}
\toprule
\textbf{Metric} & \textbf{SWARM} & \textbf{SWARM+} & \textbf{Improve.} \\
\midrule
Mean Selection       & 40.03 $\pm$ 6.41\,s   & \textbf{1.20 $\pm$ 0.04\,s} & 97.0\% \\
P95 Selection        & 85.47 $\pm$ 14.18\,s  & \textbf{1.54 $\pm$ 0.10\,s} & 98.2\% \\
P99 Selection        & 130.61 $\pm$ 99.83\,s & \textbf{2.67 $\pm$ 0.22\,s} & 98.0\% \\
Mean Sched.\ Latency & 325.22 $\pm$ 27.70\,s & \textbf{5.41 $\pm$ 0.44\,s} & 98.3\% \\
\bottomrule
\end{tabular}

\end{table}

All improvements are statistically significant ($p < 10^{-23}$, Cohen's $d > 6.4$). End-to-end scheduling latency improves from 325.22\,s to 5.41\,s ($60\times$ speedup). These gains result primarily from replacing Kafka with low-latency gRPC communication, protocol-buffer serialization, and LRU caching of repeated feasibility and cost computations (\S\ref{subsec:selection}).

The observed performance gains arise from two complementary factors: infrastructure modernization and hierarchical coordination. Under the same SWARM+ infrastructure, Hier-30 (0.93\,s) achieves a $3.0\times$ improvement over Mesh-30 (2.79\,s) with identical agent count, while Hier-110 (1.01\,s) achieves a $5.9\times$ improvement over Mesh-90 (5.95\,s) despite 22\% more agents. These gains result directly from confining consensus operations to bounded intra-group coordination domains. Conversely, Mesh-10 improves from 40.03\,s in SWARM to 1.20\,s in SWARM+ under the same multi-site deployment (Table~\ref{tab:ccgrid-comparison}), and reaches 0.34\,s in a single-site Mesh-10 deployment (Table~\ref{tab:single-site-scaling}), underscoring the impact of communication and caching optimizations independently of topology changes.

\subsection{Scalability Evaluation}
\label{subsec:scalability}

We next evaluate whether hierarchical decentralized coordination remains efficient under increasing agent counts and distributed workloads.

\textbf{Topology Scaling.}
We evaluate scalability across mesh and hierarchical deployments under increasing agent counts and workload sizes. Table~\ref{tab:single-site-scaling} summarizes the results. Mesh exhibits near-linear growth in selection time with agent count. Mesh-90 achieves only ${\sim}$51\% completion (230/450 jobs) due to $O(n^2)$ consensus conflicts, and Mesh-110 ($O(n^2)=12{,}100$ connections) enters complete consensus livelock because the three-phase protocol exceeds the 60\,s reselection timeout. Mesh topology, which represents flat (non-hierarchical) three-phase consensus as evaluated in~\cite{ccgrid25}, is therefore fundamentally unsuitable beyond approximately 30 agents.

Hierarchical topology scales significantly better relative to Mesh. Hier-30 and Hier-110 both achieve 100\% completion, with Hier-110 reaching 1.01\,s mean selection time---comparable to Mesh-30 despite a $3.7\times$ increase in agent count---because consensus is confined to intra-group meshes of $\leq$10 agents. At 250 agents, completion remains high (98.2\%) though selection time increases to 24.53\,s. Hier-990 demonstrates that the architecture can coordinate nearly 1000 agents with 98.5\% completion. All topologies achieve balanced job distribution (Jain's Fairness Index 0.44--0.59).

The selection-time increase at Hier-250 and beyond results from coordinator group size exceeding the intra-group sweet spot. With 250~agents in 25~groups, the Level~1 coordinator pool contains ${\sim}$25~agents that must reach consensus among themselves---reproducing the same $O(g^2)$ overhead that limits flat mesh at similar group sizes. Hier-990 mitigates this through a three-level hierarchy that bounds coordinator groups, but multi-level delegation still accumulates latency across levels. These results indicate that maintaining group sizes of $\leq$10 agents at every hierarchy level is critical for sub-second selection time, and that deeper hierarchies, optimized overlays~\cite{wu2024dgro}, or gossip-based higher-level coordination 
are needed to sustain low overhead beyond ${\sim}$100 agents.
\begin{table}[h]
\centering
\caption{Topology scalability (mean $\pm$ std)}
\label{tab:single-site-scaling}
\scriptsize
\setlength{\tabcolsep}{2.8pt}
\begin{tabular}{lcccccc}
\toprule
\textbf{Cfg} & \textbf{N} & \textbf{Jobs} & \textbf{Sel.\ (s)} & \textbf{P95\ (s)} & \textbf{Wait\ (s)} & \textbf{Sched.\ (s)} \\
\midrule
Mesh-10  & 10  & 100  & 0.34$\pm$0.04 & 0.58  & 3.36$\pm$0.34  & 3.73$\pm$0.34 \\
Mesh-30  & 30  & 500  & 2.79$\pm$0.08 & 3.55  & 18.60$\pm$1.00 & 21.39$\pm$1.00 \\
Mesh-90  & 90  & 450  & 5.95$\pm$0.86 & 8.67  & 70.65$\pm$7.08 & 77.08$\pm$7.57 \\
Mesh-110 & 110 & 1000 & \multicolumn{4}{c}{\textit{Consensus livelock --- 0\% completion}} \\
\midrule
Hier-30  & 30  & 500  & \textbf{0.93$\pm$0.06}  & \textbf{1.24}   & \textbf{2.68$\pm$1.72}   & \textbf{3.62$\pm$1.71} \\
Hier-110 & 110 & 1000 & \textbf{1.01$\pm$0.02}  & \textbf{1.34}   & \textbf{14.05$\pm$4.79}  & \textbf{15.05$\pm$4.78} \\
Hier-250 & 250 & 1000 & \textbf{24.53$\pm$4.62} & \textbf{80.48}  & \textbf{38.91$\pm$6.80}  & \textbf{40.29$\pm$6.75} \\
Hier-990 & 990 & 9000 & \textbf{46.12$\pm$14.33} & \textbf{208.20} & \textbf{141.24$\pm$15.79} & \textbf{144.51$\pm$15.37} \\
\bottomrule
\end{tabular}
\vspace{2pt}
\raggedright
\footnotesize
\textit{Mesh-* denotes mesh topology deployments; Hier-* denotes hierarchical topology deployments.}
\end{table}

\textbf{Weak Scaling with Pegasus Workloads.}
We perform weak scaling experiments using Pegasus workflow traces with workload size proportional to agent count ($\approx$18--20 jobs per agent). Table~\ref{tab:pegasus-weak-scaling} reports the results.

\begin{table}[h]
\centering

\caption{Weak scaling: hierarchical topology (jobs $\propto$ agents)}
\label{tab:pegasus-weak-scaling}
\footnotesize
\setlength{\tabcolsep}{3pt}
\begin{tabular}{lrrrrr}
\toprule
\textbf{Config} & \textbf{Jobs} & \textbf{Completed} & \textbf{Sel.\ (s)} & \textbf{P95\ (s)} & \textbf{Tput\ (j/s)} \\
\midrule
Hier-30  & 547  & 547 (100\%) & 1.00 $\pm$ 0.05 & 1.43 & 0.82 \\
Hier-60  & 1094 & 1094 (100\%) & 1.01 $\pm$ 0.02 & 1.42 & 1.24 \\
Hier-110 & 2188 & 2188 (100\%) & 1.09 $\pm$ 0.06 & 1.48 & 1.64 \\
\bottomrule
\end{tabular}

\end{table}

Selection time remains nearly constant across all scales (1.00--1.09\,s), confirming that hierarchical consensus overhead does not grow with agent count. All configurations achieve 100\% completion. Throughput scales from 0.82 to 1.64 jobs/s ($2.0\times$), demonstrating efficient weak scaling under Pegasus workflow traces. These results further indicate that intra-group sizes of approximately 10 agents provide an effective balance between coordination overhead and parallel execution capacity.

\textbf{Multi-Site Overhead.}
We quantify WAN coordination overhead by comparing single-site and geographically distributed deployments. Table~\ref{tab:multi-site} summarizes the resulting selection-time slowdown.
\begin{table}[h]
\centering

\caption{Multi-site vs.\ single-site mean selection time}
\label{tab:multi-site}
\begin{tabular}{lrrrr}
\toprule
\textbf{Topology} & \textbf{Jobs} & \textbf{Single\ (s)} & \textbf{Multi\ (s)} & \textbf{Slowdown} \\
\midrule
Mesh-30  & 500  & 2.79 & 5.77 & $2.07\times$ \\
Hier-30  & 500  & 0.93 & 1.19 & $1.28\times$ \\
Hier-110 & 1000 & 1.01 & 3.77 & $3.73\times$ \\
\bottomrule
\end{tabular}

\end{table}
Hier-30 demonstrates the strongest WAN resilience with only a $1.28\times$ slowdown because most coordination remains confined to intra-site groups. In contrast, Hier-110 incurs a $3.73\times$ slowdown due to additional inter-site coordination among Level~1 agents spanning all three sites. Nevertheless, hierarchical deployments remain substantially more WAN-efficient than equivalent flat mesh topologies.

\subsection{Resilience Evaluation}
\label{subsec:resilience}

We next evaluate whether decentralized coordination remains stable under distributed failures and correlated site outages. Experiments use the Hier-110 configuration (110 agents, 2188 Pegasus jobs, 3 FABRIC sites, 2 co-parent coordinators per group), providing a representative production-scale resilience scenario.

\textbf{Distributed and Correlated Failures.}
Failures are injected 90\,s into execution by terminating selected agent processes. Table~\ref{tab:hier110-resilience} summarizes completion rates and coordination latency across all scenarios.

\begin{table}[h]
\centering

\caption{Hier-110 resilience: 2188 Pegasus jobs, 3 sites}
\label{tab:hier110-resilience}
\footnotesize
\setlength{\tabcolsep}{3pt}
\begin{tabular}{lrrrr}
\toprule
\textbf{Scenario} & \textbf{Killed} & \textbf{Compl.\ \%} & \textbf{Sel.\ (s)} & \textbf{P95 Sel.\ (s)} \\
\midrule
Baseline (no failure) & 0 & 98.5 $\pm$ 0.7 & 1.16 $\pm$ 0.70 & 1.25 \\
Distributed-10 & 10 & 98.1 $\pm$ 1.6 & 0.70 $\pm$ 0.37 & 2.17 \\
Distributed-20 & $\sim$18 & 97.8 $\pm$ 1.1 & 2.45 $\pm$ 1.76 & 5.93 \\
Site outage & 11 & 95.6 $\pm$ 4.4 & 1.89 $\pm$ 1.49 & 3.49 \\
\bottomrule
\end{tabular}

\end{table}

Baseline runs achieve 98.5\% completion, confirming stable operation under normal conditions. The remaining ${\sim}$1.5\% of jobs are not lost but remain pending in the Level~1 delegation pipeline at experiment termination; once jobs reach Level~0, execution completes in all cases. The Distributed-10 case, where failure was injected to 10 agents, maintains 98.1\% completion despite 9.1\% capacity loss, while the Distributed-20 case achieves 97.8\% despite losing approximately 16\% of the agents. The increase in P95 selection latency from 1.25\,s to 5.93\,s reflects additional reselection activity required to recover orphaned jobs after failures.

Correlated site outages are more disruptive because failures simultaneously remove an entire coordination group and its associated resources. Completion decreases to 95.6\%, and mean selection time increases to 1.89\,s. Nevertheless, SWARM+ continues operating without centralized recovery services. Distributed failures are tolerated more effectively than correlated outages because partial capacity remains available across all groups, preserving quorum formation and workload feasibility.

Recovery is not free: SWARM+ autonomously reselects jobs orphaned by failed agents through the delegation monitoring mechanism described in \S\ref{subsec:topology}, triggering additional consensus rounds. Across all failure runs, the job assignment logs contain $2.0$--$2.3\times$ as many entries as unique jobs, reflecting the reselection effort required to recover orphaned work. Despite this overhead, mean selection time remains under 2.5\,s across all failure scenarios, demonstrating that delegation monitoring increases coordination volume without proportionally increasing coordination latency. 

\textbf{Coordinator Failover.}
To tolerate Level~1 coordinator failures, SWARM+ deploys $K$ co-parent coordinators per group. Table~\ref{tab:coparent-failover} evaluates this mechanism using Hier-30 with $K=2$. The lower baseline selection time (0.81\,s vs.\ 0.93\,s in Table~\ref{tab:single-site-scaling}) reflects the reduced per-group coordination load introduced by co-parent redundancy.

\begin{table}[h]
\centering

\caption{Co-parent failover: Hier-30 (500 jobs, $K=2$)}
\label{tab:coparent-failover}
\footnotesize
\setlength{\tabcolsep}{3pt}
\begin{tabular}{lrrrr}
\toprule
\textbf{Scenario} & \textbf{Completed} & \textbf{Rate} & \textbf{Sel.\ (s)} & \textbf{Sched.\ (s)} \\
\midrule
Baseline (no failure) & 500 $\pm$ 0 & 100\% & 0.81 $\pm$ 0.05 & 1.42 $\pm$ 0.06 \\
Coordinator killed ($t$=60\,s) & 491 $\pm$ 9 & 98.1\% & 0.81 $\pm$ 0.04 & 1.39 $\pm$ 0.02 \\
\bottomrule
\end{tabular}

\end{table}

Coordinator failover introduces no measurable selection-latency overhead (0.81\,s in both scenarios), demonstrating seamless decentralized redundancy. The modest completion drop (491 vs.\ 500 jobs) reflects jobs in transit to the failed coordinator at the time of failure, which are recovered through delegation monitoring but may not have completed before experiment termination.

\subsection{Impact of Data Transfer Node Awareness}
\label{subsec:dtn-impact}

SWARM+ integrates DTN connectivity as a first-class factor in its cost model (\S\ref{subsec:selection}). We compare workloads with and without DTN dependencies across both Mesh-30 and Hier-110 topologies. In DTN-aware mode, each agent is associated with 1--4 DTNs from a pool of 10 endpoints (connectivity scores in $[0.6, 0.95]$), and jobs inherit DTN dependencies from their resource profiles. The cost function applies a connectivity penalty $1 + \beta(1 - \bar{s})$, where $\bar{s}$ is the agent's mean connectivity score for the job's required DTNs and $\beta$ is the penalty factor. In DTN-unaware mode, jobs have no DTN requirements and no connectivity penalty is applied. Table~\ref{tab:dtn-impact} summarizes the results. These experiments use a single-site deployment to isolate DTN-related overhead from WAN latency effects; absolute selection times therefore differ from the multi-site results in Table~\ref{tab:single-site-scaling}.

\begin{table}[h]
\centering
\caption{Impact of DTN-aware scheduling (single-site deployment)}
\label{tab:dtn-impact}
\footnotesize
\setlength{\tabcolsep}{3pt}
\begin{tabular}{llrr}
\toprule
\textbf{Topology} & \textbf{Mode} & \textbf{Sel.\ (s)} & \textbf{P95 Sel.\ (s)} \\
\midrule
Mesh-30 & No DTN & 1.32 & 5.12 \\
(500 jobs) & DTN & 1.58 & 5.67 \\
 & \textit{$\Delta$} & \textit{+20\%} & \textit{+11\%} \\
\midrule
Hier-110 & No DTN & 1.53 & 9.47 \\
(2188 jobs) & DTN & 1.49 & 7.30 \\
 & \textit{$\Delta$} & \textit{$-$2.4\%} & \textit{$-$22.9\%} \\
\bottomrule
\end{tabular}

\end{table}

At the Mesh-30 scale, DTN-aware scheduling introduces a modest selection-time increase (1.58\,s vs.\ 1.32\,s, $+$20\%) due to connectivity penalty computation. DTN awareness narrows the candidate pool by penalizing poorly-connected agents, which slightly increases consensus overhead but produces more consistent run-to-run performance.

At hierarchical scale, DTN awareness introduces negligible mean overhead ($-$2.4\%) while delivering a 22.9\% reduction in P95 tail latency (9.47\,s to 7.30\,s). The tail improvement is most pronounced at Level-0 agents, where DTN penalties steer jobs away from poorly-connected nodes that would otherwise trigger reselection cascades. These results demonstrate that DTN-aware cost evaluation improves scheduling quality without adding measurable coordination overhead, with benefits amplified by hierarchical delegation where routing decisions at each level compound connectivity-related inefficiencies.



\subsection{Impact of Network Impairment}
\label{subsec:advanced}
We evaluate WAN sensitivity by injecting controlled impairments 
on the Hier-110 deployment (2188 Pegasus jobs, 30 VMs). Table~\ref{tab:netem-impact} summarizes the impact of packet loss and additional per-hop latency on coordination performance.

Packet loss is tolerated relatively well, with moderate degradation under 1--2\% loss. In contrast, additional latency significantly impacts decentralized consensus because the three-phase protocol compounds delay across multiple coordination rounds, eventually triggering reselection timeouts. These results identify network latency, rather than packet loss, as the dominant factor affecting decentralized consensus scheduling in geo-distributed environments.

\begin{table}[h]
\centering
\caption{Impact of network degradation on Hier-110 selection time}
\label{tab:netem-impact}
\vspace{-4pt}
\footnotesize
\setlength{\tabcolsep}{2.5pt}
\begin{tabular}{
p{1.9cm}
>{\centering\arraybackslash}p{1.7cm}
>{\centering\arraybackslash}p{0.8cm}
>{\centering\arraybackslash}p{0.8cm}
>{\centering\arraybackslash}p{0.8cm}
>{\centering\arraybackslash}p{0.8cm}
}
\toprule
\textbf{Condition} & \textbf{Completed} & \textbf{Mean\ (s)} & \textbf{Med.\ (s)} & \textbf{P95\ (s)} & \textbf{Tput\ (j/s)} \\
\midrule
Baseline
& 2188 (100\%)
& 1.09
& 1.02
& 1.48
& 1.64 \\

+25\,ms delay
& 341 (15.6\%)
& 92.40
& 10.10
& 30.61
& 0.02 \\

+50\,ms delay
& 145 (6.6\%)
& 18.07
& 15.25
& 39.39
& 0.66 \\

1\% pkt loss
& 1791 (81.9\%)
& 2.22
& 1.96
& 4.50
& 1.18 \\

2\% pkt loss
& 1899 (86.8\%)
& 5.20
& 2.81
& 11.31
& 1.20 \\

\bottomrule
\end{tabular}

\end{table}

\section{Related Work}
\label{sec:related}

Prior work related to SWARM+ spans distributed consensus, hierarchical resource management, data-aware scheduling, and resilient distributed coordination.

\textbf{Consensus and Distributed Scheduling.}
Consensus protocols such as PBFT~\cite{pbft}, Paxos~\cite{paxos}, and Raft~\cite{raft} provide distributed agreement under Byzantine or crash faults, while systems including Omega~\cite{omega}, Firmament~\cite{firmament}, Medea~\cite{medea}, Slurm~\cite{slurm}, PBS~\cite{pbs}, and HTCondor~\cite{condor} rely on centralized or shared-state scheduling architectures. Auction-based and peer-to-peer schedulers~\cite{bellagio,tycoon,p2pmpi} reduce centralization but generally provide best-effort coordination semantics. In contrast, SWARM+ combines a hierarchical three-phase consensus protocol with crash-fault-tolerant majority quorum and strongly consistent job-level workload assignment.

\textbf{Hierarchical Coordination and Data-Aware Scheduling.}
Hierarchical resource managers and federation systems~\cite{mesos,yarn,k8s-federation,gridway,gridbus,koala} improve scalability through hierarchical partitioning and multi-cluster coordination but retain centralized coordination assumptions. Data-aware scheduling systems including Hadoop~\cite{hadoop}, Spark~\cite{spark}, Quincy~\cite{quincy}, Pegasus~\cite{deelman2020jocs}, Makeflow~\cite{makeflow}, Globus~\cite{globus}, and DataMPI~\cite{datampi} optimize locality-aware placement and data movement, typically under centralized or single-domain scheduling models. SWARM+ integrates hierarchical decentralized consensus with DTN-aware workload placement and distributed locality-aware scheduling.

\textbf{Fault Tolerance and Resilience.}
Failure detection and recovery frameworks~\cite{swim,phi-accrual,hadoop,condor,paxos-reconfig,flexible-paxos,zookeeper} provide gossip-based membership management, retries, checkpointing, speculative execution, and quorum reconfiguration under failures and dynamic membership changes. However, many of these approaches still depend on centralized coordination or explicit reconfiguration mechanisms. SWARM+ instead combines decentralized failure detection, adaptive quorum management, automatic workload reselection, and coordinator failover directly within the workload management layer.

\section{Conclusions and Future Work}
\label{sec:conclusion}


In this paper, we presented SWARM+, a decentralized workload management system for geo-distributed scientific infrastructures. SWARM+ builds on our prior work on multi-agent decentralized consensus for distributed job selection to address scalability and resilience challenges. 

SWARM+ introduced three novel contributions to overcome decentralized coordination barriers: (1)~a hierarchical coordination architecture that reduces coordination overhead across distributed agents resulting in efficient consensus and improved scalability; 
(2)~decentralized resilience mechanisms that address agent failures and dynamic membership changes;
and (3)~data-aware scheduling policies that incorporate resource feasibility checks, improved cost models, network connectivity, and DTN locality.

Experimental evaluation of SWARM+ on the FABRIC testbed demonstrated scalability across up to 990 distributed agents with approximately 1\,s per-job selection time at 110 agents, balanced workload distribution, 
and reduced coordination overhead. Compared to the original SWARM system, SWARM+ improved both selection time and scheduling latency by 97--98\%. Our results showed that SWARM+ improves the system resilience, maintaining greater than 97\% job completion under distributed failures with graceful degradation under correlated site outages, and tolerating coordinator agent failures gracefully. Our results demonstrated that data-awareness in the cost function improves the scheduling quality without adding measurable coordination overhead. We also showed that increased network latencies have a negative impact on distributed consensus performance.

Future work includes gossip-based consensus and optimized overlays~\cite{wu2024dgro} for WAN-scale deployments, adaptive hierarchy construction, 
and integration with workflow management systems 
(Pegasus~\cite{deelman2020jocs}, Nextflow~\cite{nextflow}) through scheduler adapters.

\section*{Acknowledgments}
This work is supported by the U.S. Department of Energy under the Distributed Resilient Systems for Science program, grant DE-SC0024387. We acknowledge the FABRIC Testbed (NSF \#2330891).

\bibliographystyle{IEEEtran}
\bibliography{references}

\end{document}